# Nature and distribution of iron sites in a sodium silicate glass investigated by neutron diffraction and EPSR simulation

Coralie Weigel,[1] Laurent Cormier[*,1] Georges Calas,[1] Laurence Galoisy,[1] and Daniel T. Bowron[2]

[1]*Institut de Minéralogie et de Physique des Milieux Condensés, Université Pierre et Marie Curie-Paris 6, Université Denis Diderot, CNRS UMR 7590, IPGP, 4 place Jussieu, 75005 Paris, France*

[2]*ISIS Facility, CCLRC Rutherford Appleton Laboratory, Chilton, Didcot, Oxon OX11 OQX, UK*

[*]Corresponding author:    Laurent Cormier :

E-mail address: cormier@impmc.jussieu.fr

tel: +33 1 44 27 52 39

Fax : +33 1 44 27 50 32

**Abstract**

The short and medium range structure of a $NaFeSi_2O_6$ (NFS) glass has been investigated by high-resolution neutron diffraction with Fe isotopic substitution, combined with Empirical Potential Structure Refinement (EPSR) simulations. The majority (~60%) of Fe is 4-coordinated ($^{[4]}Fe$) and corresponds only to ferric iron, $Fe^{3+}$, with a distance $d_{^{[4]}Fe^{3+}-O} = 1.87 \pm 0.01 Å$. This is at variance with the 3D-structure predicted by glass stoichiometry. The existence of a majority of $^{[4]}Fe^{3+}$ sites illustrates a glass structure that differs from the structure of crystalline $NaFeSi_2O_6$, which contains only octahedral $Fe^{3+}$. The EPSR modeling of glass structure shows that $^{[4]}Fe^{3+}$ is randomly distributed in the silicate network and shares corner with silicate tetrahedra. The existence of a majority of $^{[4]}Fe^{3+}$ sites differs from the structure of the corresponding crystalline phase, which contains only octahedral $Fe^{3+}$. The network-forming behavior of $^{[4]}Fe^{3+}$, coupled with the presence of $Na^+$ ions acting as charge-compensators, is at the origin of peculiar physical properties of Fe-bearing glasses, such as the increase of the elastic modulus of sodium silicate glasses with increasing Fe-concentration. Our data provide also direct evidence for 5-coordinated Fe, with an average distance $d_{^{[5]}Fe-O} = 2.01 \pm 0.01 Å$. This second Fe population concerns both $Fe^{2+}$ and $Fe^{3+}$. 5-coordinated Fe atoms tend to segregate by sharing mainly edges. The direct structural evidence of the dual role of ferric iron in NFS glass provides support for understanding the peculiar properties of NFS glass, such as magnetic, optical, electronic or thermodynamic properties.

## 1. Introduction

Despite the fact that non-crystalline materials lack long range translational and orientational order, there is increasing evidence for some structural ordering over short and medium length scales (2-20 Å) [1]. Cations determine major structure-property relationships in glasses [2] and possess peculiar structural properties, such as unusual coordination states or heterogeneous spatial distribution [1,3-5]. However, the determination of medium-range structure in glassy systems remains a challenge, a problem exacerbated in multicomponent glasses. Diffraction experiments are widely used to get structural information on glasses, as they give access to atomic correlations at short and medium range scales. However, such information is limited because the partial functions are largely superimposed in the reciprocal and real space. Contrast techniques such as neutron diffraction with isotopic substitution (NDIS) [3,4] or the coupling between neutron and X-ray diffraction [6] can give additional structural information by separating the partial functions. However, diffraction data can only provide information on the Medium-Range Order (MRO) if they are inverted within a structural model assessing the atomic positions, and hence the MRO. Reverse Monte Carlo (RMC) and Empirical Potential Structure Refinement (EPSR) methods have been developed to adjust quantitatively diffraction data with a 3-dimensional atomistic modeling [7]. These developments make it possible to identify and characterize MRO even in compositionally complex glasses.

$Fe^{3+}$ is the most abundant redox state of iron in the majority of oxide glasses, including technological glasses, either as an unwanted impurity or as an intentionally added glass component. It is also an important component of terrestrial volcanic glasses and magmatic silicate melts. Its presence affects important properties, such as crystalline nucleation and optical and magnetic properties, as well as rheological and thermodynamic properties of the corresponding melts they are quenched from [8-10]. The study of the structural behavior of Fe is complicated under most synthesis conditions by the coexistence of two oxidation states, ferric iron, $Fe^{3+}$, and ferrous iron, $Fe^{2+}$. This limits an accurate determination of the nature and distribution of Fe-sites. Despite numerous studies, the structural behavior of $Fe^{3+}$ and $Fe^{2+}$ is still debated in silicate glasses. Most structural studies of Fe-bearing glasses have been limited to the determination of the short-range order (SRO) around Fe. $Fe^{2+}$ has been described in 6-fold coordination [11-14] but recent models suggest predominant 5-coordination within a distribution of coordination numbers from 4 to 6 [15-18]. Concerning $Fe^{3+}$, data agree with its location in 4-fold coordinated sites in alkali-bearing silicate glasses [8,11-13,19,20]. However, there is structural evidence for the presence of minor coordination states, based on diffraction [21,22] or spectroscopic data [11,16,23,24]. The presence of non-tetrahedral $Fe^{3+}$ may explain physical properties such as the rheological behavior of silicate melts or the $Fe^{3+}$ partial molar volume of silicate melts and glasses [25]. In addition, the nature of the MRO around Fe is still an open question despite optical and magnetic evidence of iron clustering in glasses [8,26-29].

We report in this work the structural study of a NFS glass, of nominal composition $NaFeSi_2O_6$, using high-resolution neutron diffraction coupled with Fe isotopic substitution and EPSR modeling. This glass composition provides a high $Fe^{3+}$ redox state (~ 88 % of total Fe), thus minimizing the influence of $Fe^{2+}$ inevitably present in silicate glasses. We show that $Fe^{3+}$ occurs mostly in tetrahedral sites ($^{[4]}Fe^{3+}$) with minority 5-coordinated $Fe^{3+}$ and $Fe^{2+}$ ($^{[5]}Fe^{3+}$ and $^{[5]}Fe^{2+}$, respectively). The existence of a majority of $^{[4]}Fe^{3+}$ sites differs from the structure of the corresponding crystalline phase, which contains only octahedral $Fe^{3+}$. The spatial distribution of Fe-sites indicates a significant clustering of $^{[5]}Fe$, at the origin of the peculiar optical and magnetic properties of Fe-bearing glasses.

## 2. Experimental section

### 2.1. Sample preparation

Two glasses were prepared from stoichiometric mixtures of dried, reagent grade $Na_2CO_3$, $SiO_2$, and $^{Nat}Fe_2O_3$ or $^{57}Fe_2O_3$ (95.86% $^{57}Fe$) for the samples labeled NFS-nat and NFS-57, respectively. Powder mixtures were decarbonated at 750°C during 12 h in platinum crucibles. Starting materials were melted at 1100°C in an electric furnace in air for 2 h. The temperature was then brought to 1300°C for 2 h and finally to 1450°C for 30 min. The melts were quenched by rapid immersion of the bottom of the crucible in water, ground to a powder and re-melted with the same cycle. This grinding-melting process was repeated three times to ensure a good chemical homogeneity.

Both glasses were dark brown and appeared bubble-free. No evidence of heterogeneity was observed during examination with an optical microscope under polarized light. An observation of the samples using transmission electron microscope (TEM) showed no secondary phases (crystalline or amorphous) at the nm scale. The effective composition was determined using electron microprobe (Table 1). The redox state, defined as the relative abundance of $Fe^{3+}$, was determined by Mössbauer spectroscopy to be $Fe^{3+}/Fe_{tot}$ = 88 ± 2 %. Glass densities were measured by Archimedes method, with toluene as a liquid reference (Table 1).

*2.2.   Isotopic substitution neutron diffraction and data corrections*

Neutron elastic diffraction experiments were performed at room temperature at the ISIS (Rutherford Appleton Laboratory, UK) spallation neutron source on the SANDALS diffractometer. The time-of-flight diffraction mode gives access to a wide Q-range: 0.3-50Å$^{-1}$. The samples were crushed and poured in a flat TiZr cell. Measurements of the samples were performed during 12 h to obtain a good signal to noise ratio. Additional measurements for shorter durations were carried out on the vacuum chamber, on the empty can and on a vanadium reference. Instrument background and scattering from the sample container were subtracted from the data. Data were merged, reduced and corrected for attenuation, multiple scattering and Placzek inelasticity effects using the Gudrun program, which is based on the codes and methods of the widely used ATLAS package [30].

The quantity measured in a neutron diffraction experiment is the total structure factor F(Q). It can be written in the Faber-Ziman formalism [31] as follows:

$$F(Q) = \sum_{\alpha,\beta=1}^{n,n} c_\alpha c_\beta \overline{b_\alpha}\,\overline{b_\beta} \left[A_{\alpha\beta}(Q) - 1\right] \qquad (1)$$

where n is the number of distinct chemical species, $A_{\alpha\beta}(r)$ are the Faber-Ziman partial structure factors, $c_\alpha$ and $c_\beta$ are the atomic concentrations of element $\alpha$ and $\beta$, and $b_\alpha$ and $b_\beta$ are the coherent neutron scattering lengths.

The total correlation function, T(r), is obtained by Fourier transforming the total structure factor F(Q). T(r) is linked to the individual distribution functions $g_{\alpha\beta}(r)$ by the weighted sum:

$$T(r) = 4\pi\rho_0 r \left( \sum_{\alpha,\beta=1}^{n,n} c_\alpha c_\beta \overline{b_\alpha}\,\overline{b_\beta} \left[g_{\alpha\beta}(r) - 1\right] + \left(\sum_i c_i b_i\right)^2 \right) \qquad (2)$$

The isotopic substitution technique consists in determining the scattering of two samples that differ only in the isotopic content of a given atom (iron) [4]. The difference between the total structure factors of NFS-nat and NFS-57 gives a first difference structure factor, $\Delta_{Fe}(Q)$:

$$\Delta_{Fe}(Q) = F_{NatFe}(Q) - F_{57Fe}(Q) = 2\sum_{\alpha \neq Fe}^{n-1} c_\alpha c_{Fe} b_\alpha (b_{Nat_{Fe}} - b_{57_{Fe}})[A_{\alpha Fe}(Q) - 1] + c_{Fe}^2 (b_{Nat_{Fe}}^2 - b_{57_{Fe}}^2)[A_{FeFe}(Q) - 1]$$

(3)

The Fourier Transform of $\Delta_{Fe}(Q)$ gives the first difference correlation function centered on iron, $T_{Fe}(r)$, characterizing the specific environment of iron. The neutron weighting factors for each atomic pair in the total structure factors and in the first difference are given in Table 2. They allow the evaluation of the different pair contributions in the scattering data.

*2.3.   EPSR simulations*

The glass structure was simulated using the EPSR code in order to extract detailed structural information about both the iron environment and the silicate network. This method allows developing a structural model for liquids or amorphous solids for which diffraction data are available. It consists in refining a starting interatomic potential by moving the atomic positions to produce the best possible agreement between the simulated and the measured structure factors [32]. A cubic box is built with the correct density, corresponding to a box size of 37.98 Å and containing 400 Fe atoms, 400 Na atoms, 800 Si atoms, and 2400 O atoms. The starting potential between atom pairs was a combination of Lennard-Jones and Coulomb potentials. The potential between atoms *a* and *b* can be represented by:

$$U_{ab}(r) = 4\varepsilon_{ab}\left[\left(\frac{\sigma_{ab}}{r}\right)^{12} - \left(\frac{\sigma_{ab}}{r}\right)^{6}\right] + \frac{1}{4\pi\varepsilon_0}\frac{q_a q_b}{r}$$

(4)

where $\varepsilon_{ab} = \sqrt{\varepsilon_a \varepsilon_b}$, $\sigma_{ab} = 0.5(\sigma_a + \sigma_b)$ and $\varepsilon_0$ is the permittivity of empty space. The Lennard-Jones $\varepsilon$ and $\sigma$ values were adjusted for NFS-nat glass until the first peak in Si-O, Fe-O and Na-O radial distribution functions is located at about 1.63 Å, 1.88 Å and 2.3 Å, respectively. The reduced depths ($\varepsilon$) and effective charges [32] were used for the reference potentials (Table 3). The simulations were run at 1000 K. They were performed in three steps to obtain the final atomic configurations. The first step consists in refining the atomic positions using only the reference potential until the energy of the simulation reaches a constant value. Then, the empirical potential refinement procedure is started: the empirical potentials are refined at the same time as the atomic positions, in order to decrease the difference between simulated F(Q) and experimental data. Once a satisfactory fit is obtained, the last step consists in accumulating simulation cycles in order to get an average information. Six distinct procedures were run for NFS-nat sample to ensure reproducibility and increase the statistics. The results presented below are averages of those different simulations.

## 3. Results

### 3.1. Structure factors

The total neutron structure factors of NFS-nat and NFS-57, F(Q) are presented in Fig. 1. Total structure factors exhibit an excellent signal-to-noise ratio up to Q = 35 Å$^{-1}$, which allows a good resolution in the real space. The structure factors are mainly affected by the isotopic substitution below 11 Å$^{-1}$. The first peak, at 1.75 Å$^{-1}$, is dependent on Fe isotopic composition and its intensity increases with the neutron scattering length of the Fe isotope. The intensity and position of the second and the third peak are different between the two total structure factors, which shows additional Fe contributions. The structure factor of NFS-nat is different from that of glassy $(Fe_2O_3)_{0.15}(Na_2O)_{0.3}(SiO_2)_{0.55}$ [21], where the first peak is split into two components and the second peak is less intense than in $S_{NFS-nat}(Q)$. As low-Q features are related to MRO, these differences indicate that the presence of excess sodium (Na/Fe > 1) modifies the medium range organization of the silicate network as compared to the charge balanced composition (Na/Fe = 1) studied here. Structural oscillations in the first difference function (Fig. 1) extend up to 20 Å$^{-1}$, which indicates a particularly well-defined local ordering around Fe.

The first peak at low Q value appears at 1.75 ± 0.02 Å$^{-1}$ in the total structure factor of the NFS glass. Its position was determined by a fit using one Gaussian component adjusted on its low Q side and laying over a horizontal background. Although its origin remains controversial [33], mainly because it cannot be assigned directly to a specific feature in the real space [34], this peak is indicative of the MRO. Its position, $Q_P$, is associated with density fluctuations over a repeat distance $D = 2\pi/Q_P$, with an uncertainty on D given by $\sigma(D) = \dfrac{2\pi\sigma(Q_P)}{Q_P^2}$ [35], where $\sigma(Q_P)$ is the uncertainty on the position of the first diffraction peak. The value of D is 3.59 ± 0.04 Å in the NFS glass. $Q_P$ shifts to 1.82 ± 0.02 Å$^{-1}$ in the first difference function, which gives a characteristic distance associated with the presence of Fe, $D_{Fe}$ = 3.45 ± 0.04 Å. Fe brings a structural ordering with a lower characteristic repeat distance than when sodium is also considered. By contrast, $Q_P$ shifts to ~1.59 Å$^{-1}$ in the total structure factor of a $(Na_2O)_{0.2}(SiO_2)_{0.8}$ glass [35]. This shift to lower scattering vector values is consistent with an enhanced separation in real space of cation-centered polyhedra in the NFS glass. This is an indication of the peculiar structural role played by Na atoms in the NFS glass, in which they act as charge compensator for tetrahedral $Fe^{3+}$ ($^{[4]}Fe^{3+}$) but also as network modifier (see below), by contrast to alkali silicate glasses in which they only act as network modifiers.

*3.2.    Total correlation functions*

The total correlation functions of NFS-nat and NFS-57, $T_{NFS-nat}(r)$ and $T_{NFS-57}(r)$, respectively, are presented with the first difference function in Fig. 2. The scattering length of $^{nat}Fe$ (b = 9.54 fm) is higher than that of $^{57}Fe$ (b = 2.66 fm), which highlights the atomic correlations implying Fe in $T_{NFS-nat}(r)$ relative to those in $T_{NFS-57}(r)$. The first maximum in $T_{NFS-nat}(r)$ and $T_{NFS-57}(r)$ is assigned to Si-O correlations. A Gaussian fit is in agreement with the presence of $SiO_4$ tetrahedra, with $d_{Si-O} = 1.63 \pm 0.01$ Å and $CN_{Si-O} = 3.9 \pm 0.1$. The Fe-O contribution at around 1.89 Å on $T_{NFS-nat}(r)$ indicates the presence of two different Fe-environments and was fitted using two Gaussian components at 1.87 Å and at 2.01 Å (Table 4 and Fig. 2). These contributions are assigned to $^{[4]}Fe^{3+}$ and to both $^{[5]}Fe^{2+}$ and $^{[5]}Fe^{3+}$, respectively [22]. The third contribution at 2.66 Å is characteristic of $d_{O-O}$ distances in $SiO_4$ tetrahedra. There is no evidence of another O-O contribution that could be assigned to $FeO_x$ polyhedra. Such a contribution is expected at around 3.1 Å, according to the mean $d_{Fe-O}$ distance in $FeO_4$ tetrahedra. The absence of this contribution indicates a Fe-site distortion in tetrahedral and higher-coordinated Fe-sites.

The features between 3 and 6 Å arise from contributions at intermediate range and cannot be unambiguously assigned to atomic pairs at this stage. However, the fourth contribution, around 3.23 Å is more intense in $T_{NFS-nat}(r)$ than in $T_{NFS-57}(r)$, and is then present in $T_{Fe}(r)$. This contribution can then be assigned to a correlation Fe-X, where X = Fe or Si (Na is unlikely due to the low weight of the Fe-Na pair and the expected large dispersion of the corresponding distance). Two contributions appear around 4.3 Å.

The first contribution, around 4.20 Å, is equally present in $T_{NFS-nat}(r)$ and in $T_{NFS-57}(r)$ and is assigned to Si-O(2) contributions, where O(2) is the oxygen second neighbor. This is consistent with the Si-O(2) distances reported in silicate glasses of similar composition [36]. The second contribution, around 4.40 Å, is more intense in $T_{NFS-nat}(r)$ than in $T_{NFS-57}(r)$, and can be assigned to a Fe-O(2) atomic correlation, as $d_{Fe-O}$ distances are larger than $d_{Si-O}$ distances.

### 3.3. EPSR modeling of short range order

EPSR simulations were performed to gain additional structural information. The good agreement between experimental and simulated structure factors can be seen for NFS-nat in Fig. 1. The EPSR-derived partial pair distribution functions (PPDFs) for X-O pairs (X=Si, Fe, Na and O) are presented in Fig. 3. The average coordination number and the distribution among the different coordination numbers (Table 5) are determined using a cut-off distance corresponding to the first minimum in the X-O PPDFs (2.35 Å, 2.67 Å, and 3.45 Å, for Si-O, Fe-O, and Na-O respectively).

Si is 4-coordinated with a minor amount of 3-coordinated Si. This indicates a narrow distribution of $d_{Si-O}$ distances, in agreement with the small value of the Debye-Waller factor obtained by a Gaussian fit of the first peak of $T_{NFS-nat}(r)$. The average value of inter-polyhedral O-Si-O bond angles is in good agreement with the ideal value of 109.4° in regular tetrahedra.

An average Fe-coordination number of ~ 4.4 and $d_{Fe-O}$ distance of 1.89 Å are obtained by EPSR. These values are in good agreement with those determined by a Gaussian fit of the correlation functions (Table 4). EPSR simulations confirm the presence of two Fe populations, $^{[4]}Fe$ and $^{[5]}Fe$, representing ~60% and ~40% of total Fe, respectively. The PPDFs of $^{[5]}Fe$-O and $^{[4]}Fe$-O are represented in Fig. 4. The interatomic distances $d_{[4]_{Fe-O}}$ and $d_{[5]_{Fe-O}}$, 1.87 Å and 2.00 Å respectively, are in agreement with those determined by Gaussian fitting of $g_{NFS\text{-}nat}(r)$. One can also notice that the first peak in the $g_{[5]_{Fe-O}}(r)$ is broader than in the $g_{[4]_{Fe-O}}(r)$, with a long tail at high r, reflecting a more distributed environment of 5-coordinated Fe. Indeed, 5-coordinated site geometry is highly flexible, with the possibility of a continuous distribution between the two extreme site geometries represented by trigonal bipyramid and square-based pyramid (Fig. 5). The small proportion of 3- and 6-coordinated Fe is assigned to a distribution of $d_{Fe-O}$ distances around the cut-off distance corresponding to 4- and 5-coordinated Fe. The average value of O-Fe-O bond angles is close to 100° and the distribution is wider than in the case of O-Si-O: this may be explained by the presence of two distinct Fe-site geometries and by a distortion of the Fe-polyhedra. The partial correlation function $g_{O\text{-}O}(r)$ presents a first maximum at 2.60 Å and a shoulder around 3.1 Å, corresponding to the contribution of the O-O linkages within the $SiO_4$ tetrahedra and $FeO_x$ (x = 4 or 5) polyhedra, respectively.

The distribution of $d_{Na\text{-}O}$ distances is not symmetric, showing a long tail at large r values. Such a distribution function is not adapted to a Gaussian fit model, which results in lower apparent coordination numbers [21]. The average coordination number of Na, obtained from EPSR simulations, is 7.0 for an average $d_{Na\text{-}O}$ distance of 2.30 Å. This distance and coordination number agree with those determined in soda lime aluminosilicate glasses [37].

*3.4.    EPSR modeling of medium range order*

The second maximum in $g_{Si\text{-}O}(r)$ and $g_{Fe\text{-}O}(r)$ appears at 4.15 Å and 4.40 Å respectively, consistent with the assignment to Si-O(2) and Fe-O(2) made on the total correlation functions. The cation-cation PPDFs can be calculated from the simulated structures and are presented in Fig. 6.

The Si-Si PPDF presents an intense first maximum at ~3.15 Å. This short distance is characteristic of corner-linked $SiO_4$ tetrahedra. The first maximum of $g_{Fe-Si}(r)$ appears around 3.33 Å. According to $d_{Si-O}$ and $d_{Fe-O}$ distances, this feature is assigned to $SiO_4$ tetrahedra sharing corners with $FeO_x$ (x = 4 or 5) polyhedra, which is confirmed by the observation of the simulated structures. The first maximum in $g_{Fe-Si}(r)$ is broader than in $g_{Si-Si}(r)$, which reflects that interpolyhedral Fe-O-Si angles are more distributed than Si-O-Si ones, partly due to the distortion of $FeO_x$ (x = 4 or 5) polyhedra. Besides, the first peak of $g_{Fe-Fe}(r)$ is asymmetric with a shoulder at low r values. It can be shown that the first maximum of $g_{^{[4]}Fe-^{[4]}Fe}(r)$ is at ~2.9 Å whereas the first maximum of $g_{^{[5]}Fe-^{[5]}Fe}(r)$ is at ~3.4 Å. According to $d_{^{[4]}Fe-O}$ and $d_{^{[5]}Fe-O}$ distances, the short $^{[5]}Fe$-$^{[5]}Fe$ distance corresponds to edge-sharing $FeO_5$ polyhedra and the longer $^{[4]}Fe$-$^{[4]}Fe$ distance to corner-sharing $FeO_4$ tetrahedra. As illustrated by the 3D-models derived from EPSR (Fig. 7), the $SiO_4$ tetrahedra are linked only by corner to the other polyhedra ($SiO_4$, $FeO_4$ et $FeO_5$), but $FeO_4$ and $FeO_5$ polyhedra share corner and/or edges with the other polyhedra, leading to the formation of tri-coordinated oxygens.

The first maximum in Na-X (X = Na, Fe or Si) PPDFs is broader than in the other PPDFs. The presence of a first maximum at ~3.2 Å in Na-Si PDDF shows that there are direct linkages between $SiO_4$ and $NaO_x$ polyhedra. In this case, Na acts as a network modifier and is associated with non-bridging oxygens. The presence of Na acting as a network modifier is not in agreement with the fully polymerized structure predicted by the glass stoichiometry, underlying the importance of the presence of 5-coordinated Fe. The first maximum in $g_{Na-Fe}(r)$, appears at ~3.30 Å and is less broad than in the case of $g_{Na-Si}(r)$, showing better defined environment of Na around Fe than around Si. This corresponds to a charge balancing behavior of Na near the negatively charged $FeO_4$ tetrahedra. The wide distribution of Na sites reflects this complex role of Na in the tetrahedral framework: it acts both as network modifier and as charge compensator.

## 4. Discussion

### 4.1. Nature of iron sites

The majority of available data indicates that $Fe^{3+}$ most commonly occurs in tetrahedral coordination in alkali and soda lime silicate glasses [10], which is in agreement with the results presented here. However, some studies have indicated the presence of another population of $Fe^{3+}$ [23,38]. This second populations has often been assigned to octahedral $Fe^{3+}$ [12,39] and recently to $^{[5]}Fe^{3+}$ [16,22,25].

Our study confirms the existence of two different Fe sites in silicate glasses. The $^{[4]}$Fe-O contribution at 1.87 Å corresponds to 60 % of total Fe [22]. The larger value of the Debye Waller factor of Fe-O vs. Si-O contributions, 0.07 Å and 0.04 Å, respectively (Table 4), shows that the FeO$_4$ tetrahedra are more distorted than the SiO$_4$ tetrahedra, with a wider distribution of $d_{Fe-O}$ distances and O-Fe-O angles. The existence of a majority of $^{[4]}Fe^{3+}$ sites differs from the structure of the corresponding crystalline phase, which contains only octahedral $Fe^{3+}$. This major structural difference explains that NaFeSi$_2$O$_6$ crystals melt incongruently into liquid and α-Fe$_2$O$_3$ at 1263 K, with a complete melting observed only at 1548 K [40].

The second Fe-O contribution at 2.01 Å (Table 4) represents $^{[5]}Fe^{2+}$ and $^{[5]}Fe^{3+}$ corresponding to ~12% and ~28% of total Fe, respectively. Despite the presence of $^{[5]}Fe^{2+}$ has not yet been demonstrated in alkali silicate glasses, it has been observed in other glass compositions [15,17,41]. The EXAFS-derived $d_{[5]Fe^{2+}-O}$ distance values of 2.00 ± 0.02 Å in alumino-silicate glasses [42] are in agreement with the experimental and calculated distances found in our study. The presence of $^{[5]}Fe^{3+}$ has been proposed recently in soda-lime silicate glasses [16] and it is consistent with $Fe^{3+}$ partial molar volume values in sodium silicate glasses [25].

5-coordinated cations are widespread in oxide glasses, despite the fact that this surrounding is unusual in the crystalline state. Such a coordination number has been shown for several important glass components, including Mg [43], Al [44], and transition elements (Ti, Fe, Ni). A detailed description of the site indicates the predominance of the square-based pyramid and trigonal bipyramid geometry for $^{[5]}Ti^{4+}$ [45] and $^{[5]}Ni^{2+}$ [46], respectively. The behavior of the partial molar volume of $Fe_2O_3$ as a function of temperature or of composition differs from that of $TiO_2$, which suggests different geometries for $Fe^{3+}$ and $Ti^{4+}$ sites [25]. From our EPSR simulations, $FeO_5$ sites correspond to a broad range of distorted polyhedra ranging from trigonal bipyramid to square-based pyramid (Fig. 5). However, the $^{[5]}Fe$ sites computed by EPSR simulations correspond to either $Fe^{3+}$ or $Fe^{2+}$ and this absence of sensitivity to the valence state can explain the wide distribution of site geometry and $d_{[5]_{Fe-O}}$ distances.

## 4.2. Iron distribution

The importance of Fe as a next nearest neighbor (NNN) of another Fe is determined by the ratio between the number of Fe NNN and the total number of its NNN's, i.e. $CN_{Fe-Fe}/(CN_{Fe-Fe}+CN_{Fe-Si})$. In a random cation distribution model, this ratio is only dependent on glass stoichiometry. For the NFS glass, a random cation distribution model predicts a value of 0.33, while EPSR simulations indicate a value of 0.42. This implies a trend for Fe to segregate. Besides, the relative importance of $^{[4]}Fe$ and $^{[5]}Fe$ in the Fe-segregation may be asserted by calculating the contribution of $^{[4]}Fe$- and $^{[5]}Fe$-NNN to the total number of NNN.

A random distribution of $^{[4]}Fe$ corresponds to a value of the ratio $CN_{[4]Fe\_[4]Fe}/(CN_{[4]Fe\_[4]Fe} + CN_{[4]Fe\_[5]Fe} + CN_{[4]Fe-Si}) = 0.20$. EPSR simulations give a ratio of 0.23 for $^{[4]}Fe$. Consequently, $^{[4]}Fe$ is randomly distributed within the glassy framework and shares only corners with other tetrahedra ($FeO_4$ or $SiO_4$) (Fig. 7). This contribution corresponds to the maximum observed in the Fe-Si PPDF and to the second contribution around 3.4 Å in the Fe-Fe PPDF (Fig. 6). The well-defined shape of this contribution of $^{[4]}Fe$ to the partials would indicate a small distribution of the $^{[4]}Fe$-O-T angles in the glassy framework.

A random distribution of $^{[5]}Fe$ corresponds to a value of the ratio $CN_{[5]Fe\_[5]Fe}/(CN_{[5]Fe\_[5]Fe} + CN_{[5]Fe\_[4]Fe} + CN_{[5]Fe-Si}) = 0.13$. EPSR simulations give instead a value of 0.24. This shows that, contrary to $^{[4]}Fe$, $^{[5]}Fe$ tends to segregate. It is important to note that TEM observations do not detect chemical or crystalline heterogeneities in the NFS glass. EPSR simulations suggest that segregated $FeO_5$ polyhedra tend to favor mutual edge-linkages (Fig. 7), corresponding to Fe-Fe contributions at shorter distances relative to a corner-sharing linkage. Such a contribution is apparent around 2.9 Å in the Fe-Fe PPDF of the EPSR model. An average number of 5 $FeO_5$ polyhedra is implied in this clustering, ranging from 2 to 10 $FeO_5$ polyhedra. Adjacent $FeO_4$ and $FeO_5$ polyhedra share both edges and corners, which leads to a wide distribution of $^{[4]}Fe$-$^{[5]}Fe$ distances and explains the broadening of the first peak of the Fe-Fe PPDF.

Non-uniform concentration of $Fe^{3+}$ ions in the glass can give rise to peculiar Mössbauer spectra, with regions of larger and lower $Fe^{3+}$ concentration giving rise to a doublet and to a resolved sextet, respectively [29]. A trend for clustering was also detected by a spectroscopic investigation of silicate glasses with $Fe_2O_3$ > 1 mol%, which was interpreted as resulting from the formation of Fe clusters, including $Fe^{2+}$-O-$Fe^{3+}$ and $Fe^{3+}$-O-$Fe^{3+}$ interactions [8]. Peculiar magnetic properties, such as the antiferromagnetism shown by electron paramagnetic resonance spectra of concentrated glasses [47], arise from short $d_{Fe^{3+}-Fe^{3+}}$ distances. The short $d_{[5]Fe-[5]Fe}$ distance observed at 2.9 Å in the EPSR PPDFs describes edge sharing Fe-polyhedra and hence should give rise to magnetic properties to the NFS glass. On the other hand, energetically favorable interaction between $Fe^{2+}$ and $Fe^{3+}$ should cause clustering of mixed-valence species [48].

*4.3. Structural significance of iron sites*

The majority of $Fe^{3+}$ is 4-coordinated with an average $d_{[4]Fe^{3+}-O} = 1.87 \pm 0.01 Å$. The presence of 4-coordinated $Fe^{3+}$ has been shown in sodium silicate glasses and its presence is often associated to a network forming behavior [12,13]. In the NFS glass, 4-coordinated $Fe^{3+}$ has 3.1 Si and 1.2 [4]Fe neighbor. Those EPSR-derived coordination numbers are consistent with [4]$Fe^{3+}$ contributing to the polymerization of the silicate network. Moreover, the random distribution of [4]$Fe^{3+}$ and the corner sharing of $FeO_4$ with $SiO_4$ and $FeO_4$ are in agreement with [4]$Fe^{3+}$ acting as a glass network former. Such a behavior is expected to decrease the tendency of glass phase separation, by forming bonding with Na, similar to the effect of Al on depressing the immiscibility of sodium silicate glasses [49]. This network-forming position explains also the increase of the elastic modulus as the $Fe^{3+}$ content increases [50].

The presence of about 28% of oxygen atoms in a non-bridging position is a consequence of the presence of [5]Fe. The nominal composition of NFS glass predicts a fully polymerized glassy network. The presence of 5-coordinated Fe is then responsible for the depolymerization of the glassy network. In the EPSR model, $FeO_5$ polyhedra often share edges with other $FeO_5$ polyhedra and there is a trend for segregation of [5]Fe. The existence of glass regions enriched in [5]Fe is consistent with the modified random network model [51], where modifying cations disrupt the polymerized network and tend to segregate. In this picture, [5]Fe corresponds to a network modifying cation. The presence of [5]Fe$^{3+}$ was suspected to be at the origin of the anomalous macroscopic thermodynamic properties of the sodium iron silicate liquids and glasses, as shown by the chemical and temperature dependence of their configurational heat capacity [52]. Our study confirms this model.

As glass structure is often suspected to be that of a supercooled liquid frozen in at the glass transition temperature [1], the presence of [5]Fe sites in NFS glass may inherit some structural properties related to melt dynamics. High-temperature $^{29}$Si and $^{17}$O exchange nuclear magnetic resonance (NMR) spectroscopic studies have shown the importance of Si–O bond breaking and chemical exchange among polyanionic units in alkali silicate liquids [53]. Such atomic motions, at the origin of the viscous flow, are favored by transient unusual coordination states, such as 5-coordinated sites [53]. If we extend this picture to cations such as iron, [5]Fe is also expected as an intermediate species during viscous flow, with a higher probability than [5]Si, due to the ease of Fe-O bond breaking relative to Si-O bonds.

*4.4.    Influence of iron clustering on physical properties*

The clustering associated to the presence of [5]Fe is important to understand several glass properties. A noticeable consequence of the presence of Fe in silicate glasses is their color. At low Fe$^{3+}$-concentration, oxide glasses exhibit a light coloration, as d-d crystal field transitions are spin forbidden in a d$^5$ configuration. However, the dark brown color of the NFS glass indicates an efficient coloration mechanism. An intense O$^{2-}$-Fe$^{3+}$ charge transfer process is known in glasses, giving rise to an intense absorption band with a maximum in the UV region and a tail extending at lower wavenumbers [54]. In addition, glasses also exhibit at high Fe-concentration a $Fe^{2+}$-$Fe^{3+}$ intervalence charge transfer, which has been suspected to be due to the presence of clusters in concentrated glasses [27]. Such a clustering is shown by EPSR simulations in the NFS glass and may be observed in other concentrated Fe-bearing glasses, with [5]Fe playing a major role due to the efficiency of charge transfer processes in edge-sharing polyhedra [55].

At high concentration, iron-containing glasses behave as amorphous semiconductors [56,57]. Mixed valence conduction in these glasses is described in terms of electron-hopping between neighboring $Fe^{3+}$ and $Fe^{2+}$ [56,58]. This is consistent with the presence of mixed valence clusters, which favor electron hopping and provide pathways for charge transport. As electron hopping is made easier if the two sites have similar symmetries [27], the structure of these clusters is thus consistent with a similar 5-coordinated site for $Fe^{3+}$ and $Fe^{2+}$. Therefore both SRO and MRO are important to understand electronic conduction processes.

Sodium-silicate glasses are known to prone microphase separation [59,60]. Iron has an influence on the critical temperature, $T_c$, defined as the maximum temperature at which the phase separation is observed. The effect of iron redox, $Fe^{2+}/Fe_{tot}$, has been reported in sodium silicate [60] and borosilicate glasses [26]. With $Fe_2O_3$ doping, there is a $T_c$ downshift for glasses containing $Fe^{3+}$ ions compared to sodium silicate glasses. This result agrees with [4]$Fe^{3+}$ acting as a network former requiring sodium as charge compensator and thus acting for enhancing glass structural homogeneity, a behavior similar to Al. In oxidized glasses, $T_c$ has been shown to strongly decrease with increasing $Fe^{2+}/Fe_{tot}$ from ~ 5 to 10 % [26] and then to slightly increase at higher $Fe^{2+}$ content [60]. The initial lowering of the critical temperature was interpreted as the formation of dissimilar $Fe^{3+}$-$Fe^{2+}$ pairs while, at high $Fe^{2+}$ content, $Fe^{3+}$-$Fe^{3+}$ and $Fe^{2+}$-$Fe^{2+}$ clustering is more likely.

As a general rule, the diffusion of modifier and intermediate cations is faster than that of network formers and controls the structure and composition of nucleating phases [61]. As the $d_{[5]Fe-O}$ distance is longer than the $d_{[4]Fe-O}$ one, the $[5]$Fe-O bond strength is weaker than that of $[4]$Fe-O. $[5]$Fe should then exhibit a higher mobility than $[4]$Fe. Since $[4]$Fe is not usually found in crystals, this network forming position will prevent devitrification. Conversely, $[5]$Fe that segregates and acts as a network modifier, is the second more mobile species after Na in NFS. It is then expected that the first nucleating phases will be Fe-rich phases, which is indeed observed during the devitrification of Fe-bearing glasses. In particular, one can observe the formation of hematite $Fe_2O_3$ [62], in which Fe is 6-coordinated. Hematite is also formed during the incongruent melting of the NFS glass [40]. The regions enriched in iron may then play an important role in nucleation mechanisms in glasses and, more generally, in their thermal stability.

## 5. Conclusions

High-resolution neutron diffraction data with isotopic substitution, combined with EPSR simulations, has been used to obtain structural information on the short and medium range order around Fe in a $NaFeSi_2O_6$ glass. Three Fe-populations exist, with the evidence of $[5]Fe^{3+}$ and $[5]Fe^{2+}$ in addition to the network-forming $[4]Fe^{3+}$ predicted by glass stoichiometry. The local structure of NFS glass differs from the structure of the corresponding crystalline phase, which contains only octahedral $Fe^{3+}$. The EPSR model of the neutron data shows that $[4]Fe^{3+}$ is randomly distributed in the network and shares corners with the other cationic polyhedra, acting as a network former with Na as a charge compensator, with some structural similarity with alumino-silicate glasses. By contrast, $[5]Fe^{3+}$ and $[5]Fe^{2+}$ provide an original picture of cation ordering in silicate glasses, through an efficient clustering with edge-sharing polyhedra. The presence of these sites does not obey the structural picture of a glass, expected to be 3D according to its stoichiometry. The direct structural evidence of the dual role of ferric iron in NFS glass confirms conclusions from some earlier spectroscopic studies, which predicted that some $Fe^{3+}$ ions were not in a network-forming position in sodium silicate glasses. It also provides support for understanding the evolution of melt/glass structure and rheological properties as a function of the Fe-redox state [14]. The non-homogeneous distribution of Fe leads in NFS glass to the formation of Fe-clusters, at the origin of the peculiar properties of NFS glass, such as magnetic, optical, electronic or thermodynamic properties. The determination of the electronic structure of these amorphous clusters and their chemical stability would provide clues to a modelling of the physical properties of concentrated Fe-bearing glasses.

**ACKNOWLEDGEMENTS**

We acknowledge the support of the INSU/CNRS for funding the $^{57}$Fe isotopes. This is IPGP contribution #xxx. The authors would like to acknowledge Catherine McCammon (Bayerisches Geoinstitut, Germany) and Stéphanie Rossano (Laboratoire Géomatériaux et Géologie de l'Ingénieur, France) for Mössbauer spectroscopy and Nicolas Menguy (IMPMC, Paris) for the TEM studies.

**FIGURE CAPTION**

Fig. 1. (Color online) From top to bottom: The solid lines represent experimental total structure factors for NFS-nat, NFS-57 and first difference function. The dotted line superimposed to $S_{NFS-nat}(Q)$ represents the fit to the data obtained after EPSR modeling for NFS-nat (the feedback factor was taken equal to 0.75). The curves have been displaced vertically for clarity.

Fig. 2. (Color online) Experimental total correlation functions obtained by Fourier Transform (F.T.) of the total structure factors (dots). The F.T. were performed by multiplying F(Q) by a Lorch function to reduce truncation effects over the Q interval of 0.4-35 Å$^{-1}$ for total correlation functions $T_{NFS-nat}(r)$ and $T_{NFS-57}(r)$ and 0-20 Å$^{-1}$ for first difference function. The Gaussian fitting of the data is represented for $T_{Fe}(r)$ and $T_{NFS-nat}(r)$ (grey solid lines).

Fig. 3. (Color online) Cation-oxygen partial pair distribution functions extracted from EPSR structural simulations. Curves have been displaced vertically for clarity.

Fig. 4. Fe-O partial pair distribution functions extracted from EPSR structural simulations for 4- and 5-coordinated Fe. The green line, $g_{Fe-O}(r)$, represents the sum of $g_{[4]Fe-O}(r)$ and $g_{[5]Fe-O}(r)$.

Fig. 5. (Color online) Three examples of $^{[5]}$Fe site geometries: trigonal bipyramid (right), square-based pyramid (left) and an intermediate, distorted site (center).

Fig. 6. (Color online) Cation-cation partial correlation functions extracted from EPSR structural simulations. Curves have been displaced vertically for clarity.

Fig. 7. (Color online) Typical representations of the medium range organization around Fe in tetrahedral (left) and 5-coordinated sites (right), as derived from EPSR simulations.

# TABLES

Table 1

Experimental glass composition in atomic % obtained by electron microprobe analysis, atomic number densities (at. Å$^{-3}$).

| Sample | Fe ± 0.1 % | Si ± 0.2 % | Na ± 0.1 % | O ± 0.2 % | d (at Å$^{-3}$) ± 0.001 at. Å$^{-3}$ |
|---|---|---|---|---|---|
| NFS-nat | 10.3 | 20.4 | 9.7 | 59.6 | 0.072 |
| NFS-57 | 10.0 | 19.7 | 10.5 | 59.8 | 0.073 |
| Theory | 10.0 | 20.0 | 10.0 | 60.0 | - |

Table 2

Neutron weighting factors for each atomic pair in the total structure factors of samples NFS-nat and NFS-57 (eq. 1) and in the first difference function $\Delta_{Fe}(Q)$ (eq. 3).

|  | Fe-Fe | Fe-Si | Fe-O | Fe-Na | Si-Si | Si-O | Si-Na | O-O | Na-O | Na-Na |
|---|---|---|---|---|---|---|---|---|---|---|
| NFS-nat | 0.0097 | 0.0167 | 0.0680 | 0.0068 | 0.0072 | 0.0586 | 0.0059 | 0.1196 | 0.0240 | 0.0012 |
| NFS-57 | 0.0007 | 0.0043 | 0.0184 | 0.0020 | 0.0067 | 0.0568 | 0.0062 | 0.1204 | 0.0261 | 0.0014 |
| $\Delta_{Fe}$ | 0.0089 | 0.0123 | 0.0495 | 0.0048 | 0.0005 | 0.0018 | -0.0003 | -0.0008 | -0.0021 | -0.0002 |

Table 3

Parameters for the starting potential in the EPSR simulations.

|  | Coulomb charges | ε (kJ/mole) | σ (Å) |
|---|---|---|---|
| $Na^+$ | +0.5 e | 0.1750 | 2.10 |
| $Fe^{3+}$ | +1.5 e | 0.1500 | 1.70 |
| $Si^{4+}$ | +2.0 e | 0.1750 | 1.06 |
| $O^{2-}$ | -1.0 e | 0.1625 | 3.60 |

Table 4

Parameters obtained by Gaussian fit of the first peak of the correlation function of NFS-nat and of the first difference function. Estimated uncertainties: d ± 0.01 Å, CN ± 0.1, and σ ± 0.01 Å.

|  | Si-O | | | Fe-O | | |
|---|---|---|---|---|---|---|
|  | d(Å) | CN | σ(Å) | d(Å) | CN | σ(Å) |
| $T_{NFS-nat}(r)$ | 1.63 | 3.9 | 0.04 | 1.87 | 3.2 | 0.07 |
|  |  |  |  | 2.01 | 1.0 | 0.07 |
| $T_{Fe}(r)$ | - | - | - | 1.89 | 4.3 | 0.09 |

Table 5

Average coordination numbers obtained using EPSR, and distribution of each coordination number for each species. The $d_{cation-O}$ distance is the position of the first peak in the X-O PPDF's. The last line is the standard deviation in the iron coordination number over six EPSR simulations.

| Atomic pair | $d_{cation-O}$ (Å) | average coord. | % 3-coord | % 4-coord | % 5-coord | % 6-coord | % 7-coord | % 8-coord | % 9-coord | % 10-coord. |
|---|---|---|---|---|---|---|---|---|---|---|
| Si-O | 1.61 | 4.0 | 0.4 | 99.5 | 0.1 | 0 | 0 | 0 | 0 | 0 |
| Fe-O | 1.89 | 4.4 | 1 | 59 | 36 | 4 | 0 | 0 | 0 | 0 |
| Na-O | 2.30 | 7.0 | 0 | 2 | 9 | 23 | 29 | 24 | 10 | 3 |
| $\Delta CN_{Fe-O}$ | - | 0.03 | 0.34 | 2.29 | 2.25 | 0.91 | 0.00 | 0.00 | 0.00 | 0.00 |

**FIGURE 1**

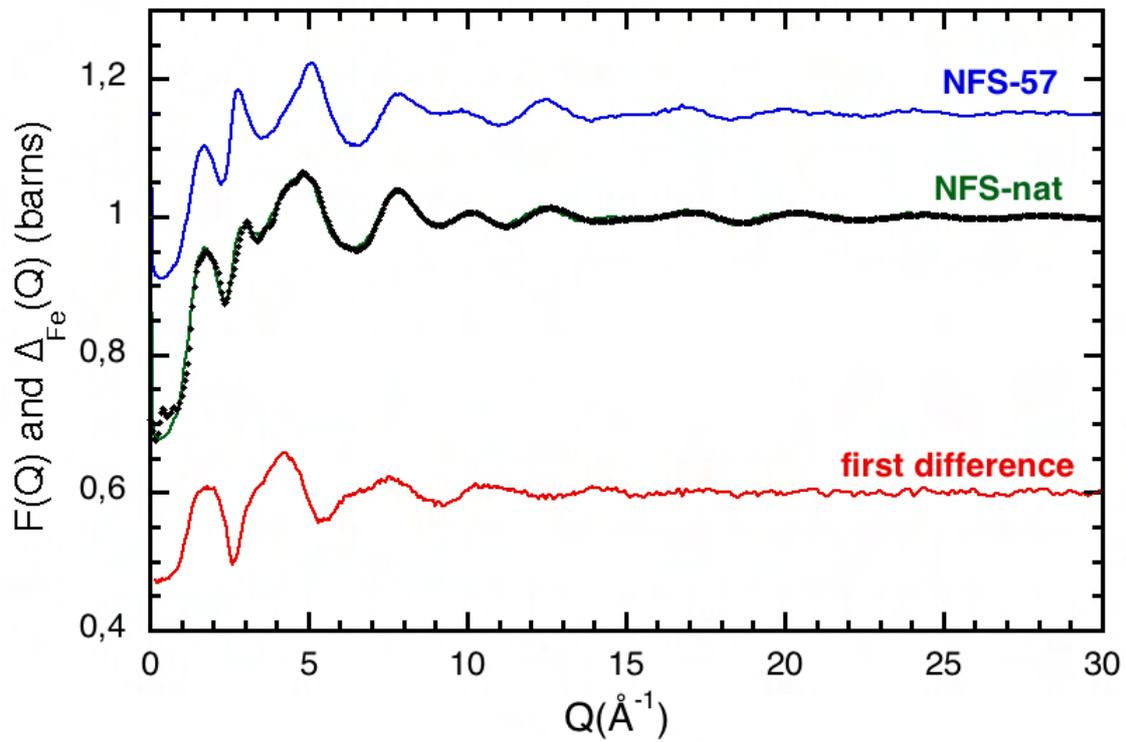

Fig. 1. (Color online) From top to bottom: The solid lines represent experimental total structure factors for NFS-nat, NFS-57 and first difference function. The dotted line superimposed to $S_{NFS-nat}(Q)$ represents the fit to the data obtained after EPSR modeling for NFS-nat (the feedback factor was taken equal to 0.75). The curves have been displaced vertically for clarity.

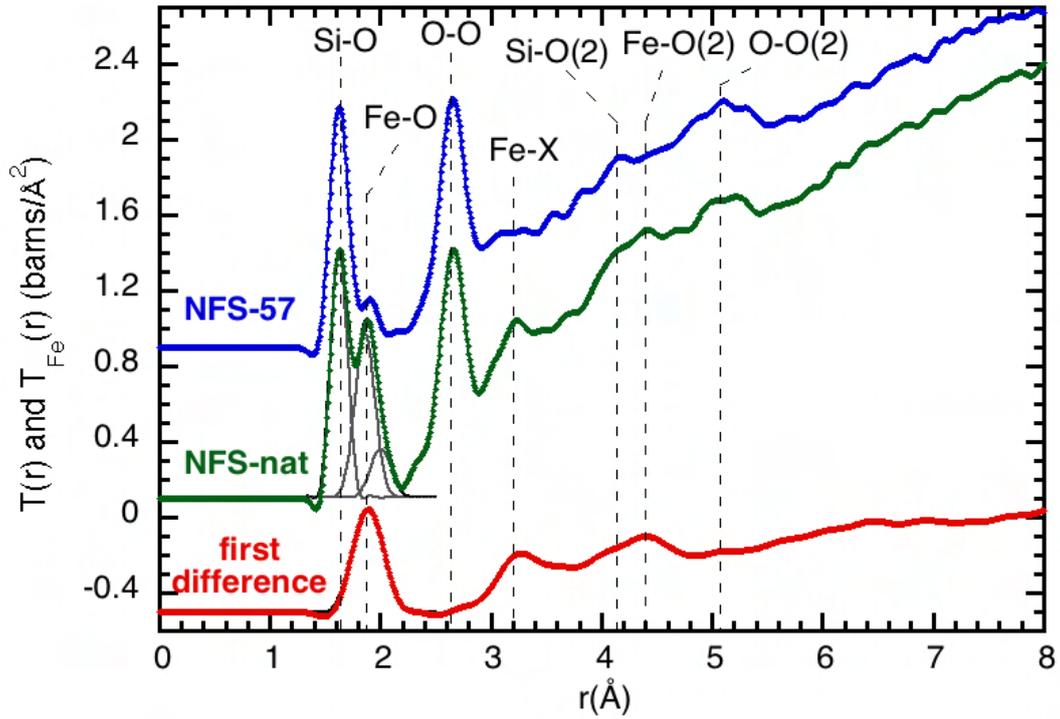

Fig. 2. (Color online) Experimental total correlation functions obtained by Fourier Transform (F.T.) of the total structure factors (dots). The F.T. were performed by multiplying F(Q) by a Lorch function to reduce truncation effects over the Q interval of 0.4-35 $\text{Å}^{-1}$ for total correlation functions $T_{NFS-nat}(r)$ and $T_{NFS-57}(r)$ and 0-20 $\text{Å}^{-1}$ for first difference function. The Gaussian fitting of the data is represented for $T_{Fe}(r)$ and $T_{NFS-nat}(r)$ (grey solid lines).

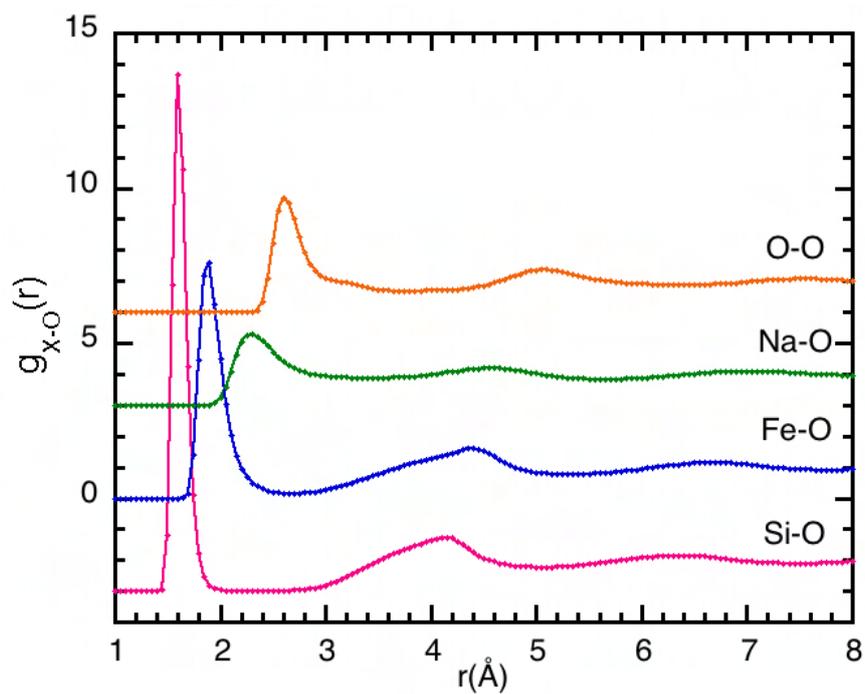

Fig. 3. (Color online) Cation-oxygen partial pair distribution functions extracted from EPSR structural simulations. Curves have been displaced vertically for clarity.

**FIGURE 4**

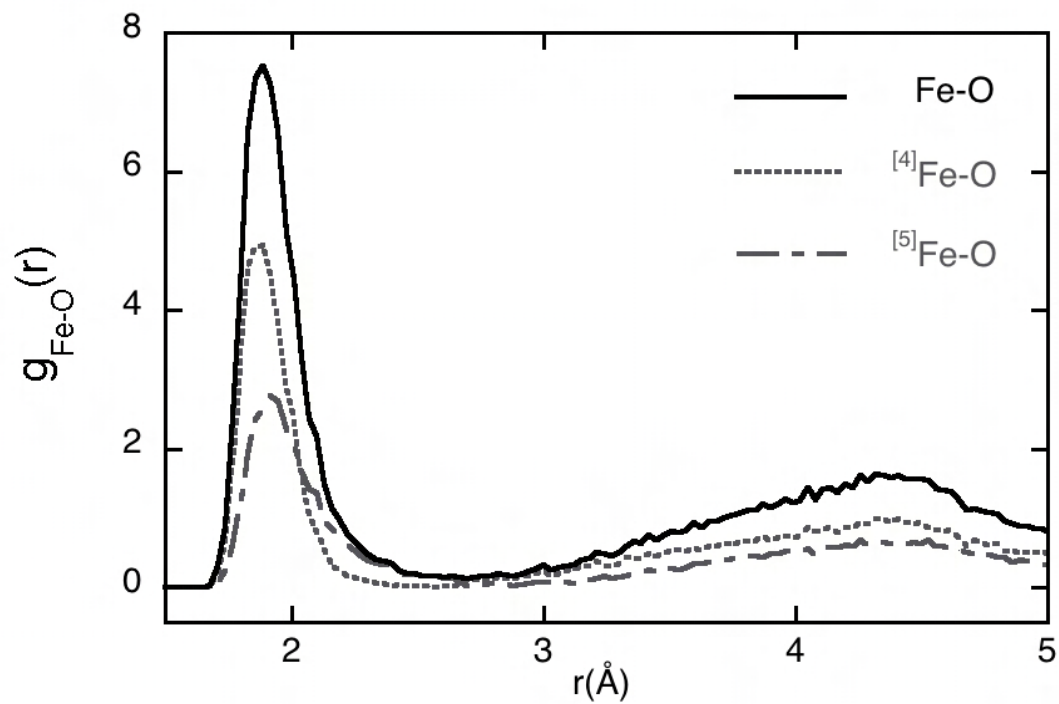

Fig. 4. Fe-O partial pair distribution functions extracted from EPSR structural simulations for 4- and 5-coordinated Fe. The green line, $g_{Fe-O}(r)$, represents the sum of $g_{[4]Fe-O}(r)$ and $g_{[5]Fe-O}(r)$.

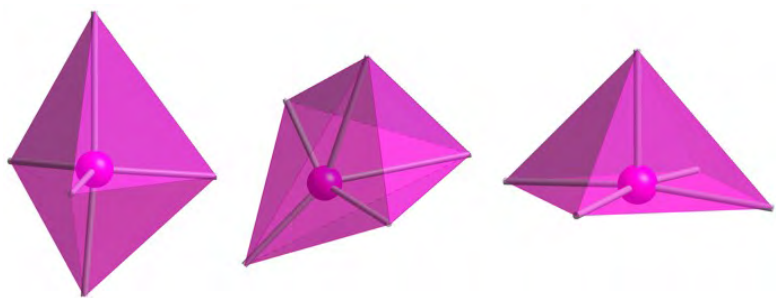

Fig. 5. (Color online) Three examples of [5]Fe site geometries: trigonal bipyramid (right), square-based pyramid (left) and an intermediate, distorted site (center).

**FIGURE 6**

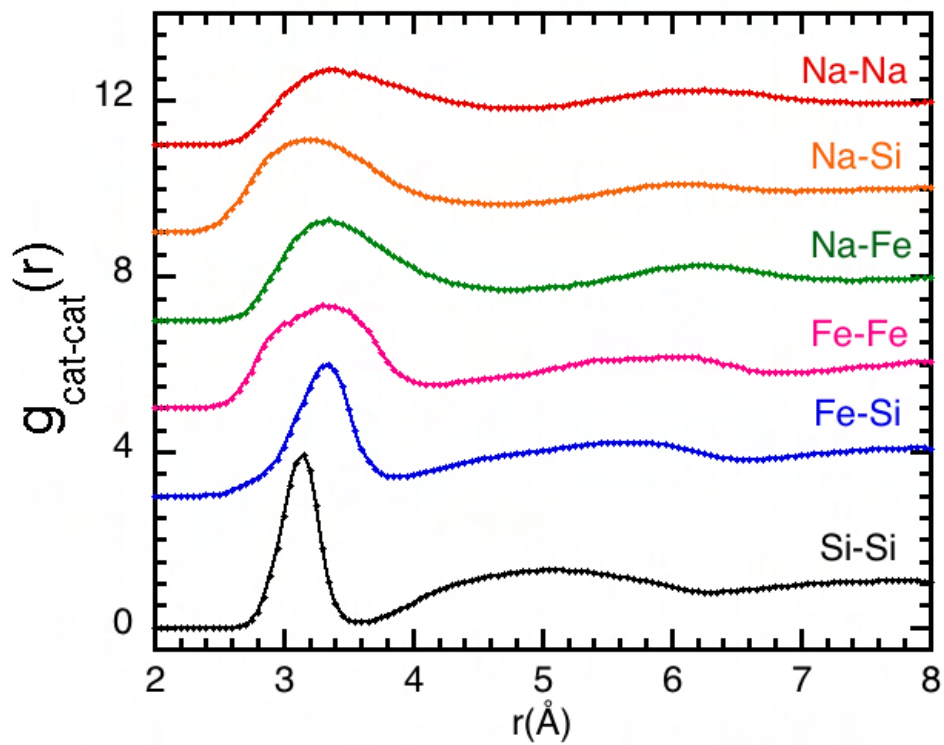

Fig. 6. (Color online) Cation-cation partial correlation functions extracted from EPSR structural simulations. Curves have been displaced vertically for clarity.

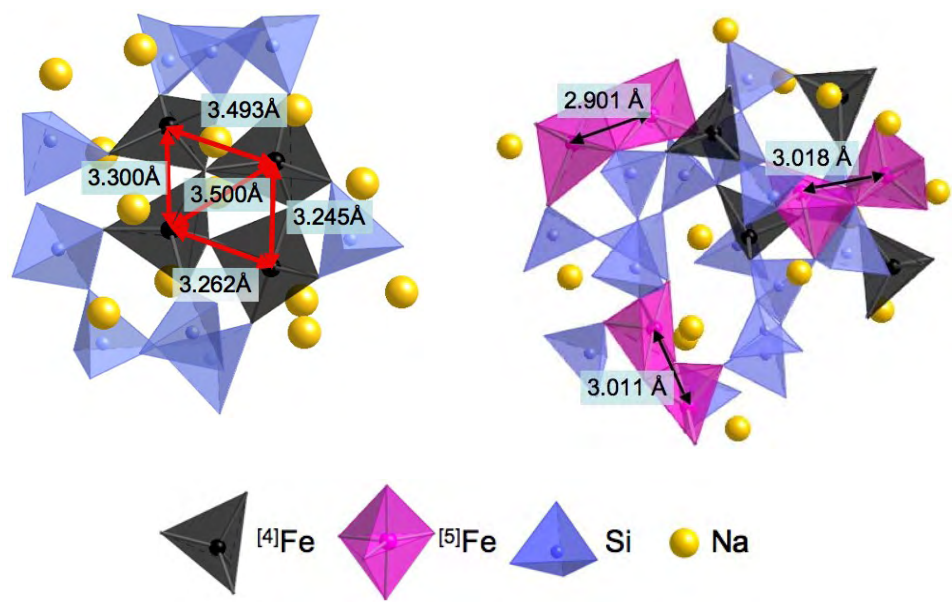

Fig. 7. (Color online) Typical representations of the medium range organization around Fe in tetrahedral (left) and 5-coordinated sites (right), as derived from EPSR simulations.